\documentclass[10pt]{article}
\usepackage{amsfonts}
\usepackage{amssymb}
\usepackage{amsthm}
\usepackage{amsmath}
\usepackage{authblk} 
\usepackage{graphicx}
\usepackage{hyperref}
\usepackage{enumerate}
\usepackage{tikz}
\usetikzlibrary{decorations.pathmorphing, patterns,shapes}
\usepackage{color}
\usepackage[mathscr]{euscript}

\usepackage{soul,todonotes}

\usepackage{mathrsfs}
\usepackage{hyperref}
\usepackage{datetime}

\numberwithin{equation}{section}

\newcommand{\ii}{{\rm i}}
\newcommand{\ee}{{\rm e}}
\newcommand{\x}{{\rm x}}

\newcommand{\dd}{{\rm d}}

\newcommand{\beq}{\begin{equation}}
\newcommand{\ene}{\end{equation}}

\newtheorem{thm}{Theorem}

\newtheorem{prop}[thm]{Proposition}

\newtheorem{conj}[thm]{Conjecture}

\theoremstyle{definition}
\newtheorem{defn}[thm]{Definition}
\newtheorem{rem}[thm]{Remark}

\newdateformat{daymonthyear}{\THEDAY \, \monthname[\THEMONTH] \THEYEAR}

\newcommand{\Af}{\mathscr{A}}
\newcommand{\Bf}{\mathscr{B}}

\begin{document}

\title{Quantum strong cosmic censorship and black hole evaporation}
\author{Benito A. Ju\'arez-Aubry}
\affil{ Department of Mathematics, University of York, Heslington, York YO10 5DD, UK\thanks{Present address}}
\affil{Instituto de Ciencias Nucleares, Universidad Nacional Aut\'onoma de M\'exico, A. Postal 20-126, CDMX, Mexico}
\affil{benito.juarezaubry@york.ac.uk}
\date{V3: \daymonthyear\today \\ Published in: Class.Quant.Grav. {\bf 41} (2024) 19, 195027 \\ DOI: \href{https://doi.org/10.1088/1361-6382/ad756c}{10.1088/1361-6382/ad756c}}

\maketitle

\begin{abstract}
It is common folklore that semiclassical arguments suggest that, in black hole evaporation, an initially pure state can become mixed. This is known as the \emph{information loss puzzle} (or {\emph paradox}). Here we argue that, if taken at face value, semiclassical gravity suggests the formation of a final singularity instead of information loss. A quantum strong cosmic censorship conjecture, for which we give a rigorous statement, supports this conclusion. Thus, there are no reasons to expect a failure of unitarity in black hole evaporation or for any quantum gravity theory that can `cure' singularities.
\end{abstract}

\section{Introduction}
\label{sec:Intro}

In 1975 Hawking made the remarkable prediction that, in the process of gravitational collapse leading to the formation of a black hole, test quantum fields will emit radiation, which at late times has a temperature proportional to the black hole surface gravity, $\kappa$, the Hawking temperature, $T_{\rm H} = \kappa/(2 \pi)$ \cite{Hawking:1975}. Thus, stationary observers at infinity detect late-time quantum radiation at the Hawking temperature. See e.g. \cite{Juarez-Aubry:2018} for an analytical calculation including time estimates in lower dimensions.

\begin{figure}
\begin{center}
\begin{tikzpicture}
\draw[thick] (-2,-2) -- (-2, 2);
\draw[thick] (-2,-2) -- (1.75, 1.75);
\draw[decoration = {zigzag,segment length = 2mm, amplitude = 0.5mm}, decorate] (-2,2) -- (-0.5,2);
\draw [thick, draw=none, fill=blue, fill opacity=0.2]
       (-2,2) -- (-0.5,2) -- (-2,0.5) -- cycle; 
\draw [thick, draw=none, pattern=north west lines, pattern color = black]
       (-2,2) -- (-1.2,2) to[out=-65,in=55] (-2,-2) -- cycle;  
\draw[thick]  (-1.2,2) to[out=-65,in=55] (-2,-2); 
\draw[thick] (-0.5,2) -- (-0.5, 4);
\draw[thick] (-0.5, 4) -- (1.75,1.75);
\draw[thick] (-0.5,2) -- (-2, 0.5);
\draw [thick, draw=none, pattern=crosshatch dots, pattern color=red, fill opacity=0.5]
       (-0.5,2) -- (-0.5,4) -- (0.5,3) -- cycle; 
\node at (-1.8,-2.1) {$i^-$};
\node at (-0.2,4.2) {$i^+$};
\node at (2,1.75) {$i^0$};
\node at (1,3) {$\mathscr{I}^+$};
\node at (0.5,-0) {$\mathscr{I}^-$};
\node at (-2.5, 0) {$r = 0$};
\node at (-1.2, 2.2) {$r \to 0$};
\node at (-1, 3) {$r = 0$};
\node at (0.3, 2.4) {$\mathscr{C}$};
\end{tikzpicture}
\end{center}
\caption{Hawking's black hole evaporation conformal diagram: Infalling matter forms a black hole region (shaded in blue). After the black hole Hawking-radiates away, a post-evaporation region emerges, which is nearly flat (denoted by red dots). A Cauchy horizon, $\mathscr{C}$, appears as the past boundary of the post-evaporation region. At the end of evaporation, a pure `in' state $\omega_{\rm in}$ defined on the observable algebra associated to $\mathscr{I}^-$ will evolve into a mixed `out' state $\omega_{\rm out}$ defined on the observable algebra associated to $\mathscr{I}^+$.}
\label{Fig:Hawking-Diagram}
\end{figure}
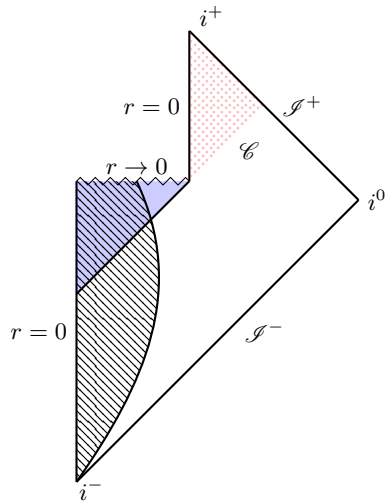

Hawking also deduced in \cite{Hawking:1975} by semiclassical arguments that the emitted radiation will lead to an eventual black hole evaporation.  
The result of evaporation -- it is argued -- is a flat or nearly-flat geometry with the quantum state of matter in a mixed state of left-over quantum radiation. However, this radiation depends only on the geometric properties of the black hole, fully characterised by its mass, charge and angular momentum, and not on the details of the matter that initially formed the black hole or went inside it. See Fig. \ref{Fig:Hawking-Diagram} for details.

In the black hole evaporation process represented in Fig. \ref{Fig:Hawking-Diagram}, an initial pure in-state at $\mathscr{I}^-$  (for example, a coherent in-state `peaked' around a classical configuration of infalling matter that forms the black hole) is unitarily inequivalent to the final out-state at $\mathscr{I}^+$, which is necessarily mixed, since $\mathscr{I}^+$ is not a Cauchy surface of the pre-evaporation region, as has been argued many times in the past (see e.g. \cite{Unruh-Wald}). 

This is the \emph{black hole information loss puzzle}, succinctly stated as the situation that in the semiclassical evaporation picture \emph{an initially pre-evaporation pure state can evolve into a post-evaporation mixed state}. Thus, quantum determinism seems to fail (loosely referred to as information loss -- and we shall continue to use this terminology here). 

There have been multiple approaches to mitigate or resolve the puzzle, which are however non-conclusive. See e.g. \cite{Kay:2022wpn, Unruh-Wald, Hossenfelder:2009xq, Preskill:1992tc} for some interesting views and historical accounts.

Our purpose here is to argue that, contrary to the usual folklore, standard semiclassical arguments do not lead to the loss of information. 
Instead, there is strong evidence of a \emph{quantum strong cosmic censorship} seemingly preventing a \emph{bona fide} semiclassical description of the final stage of evaporation. Further, we argue that, if taken at face value, semiclassical gravity suggests the formation of a final singularity instead of the Cauchy horizon of Fig. \ref{Fig:Hawking-Diagram} and no post-evaporation region where information is lost. This indicates that the endpoint of evaporation lies fully in the regime of quantum gravity, in analogy with the way in which general relativity indicates that black hole singularities lie in the regime of quantum gravity.

To make the argument precise, we mathematically state this quantum strong cosmic censorship conjecture in Sec. \ref{sec:QSCC}. In words, it states that, given a pure Hadamard state defined on the observable algebra in the interior of the domain of dependence of some non-Cauchy achronal surface, $S$, it is typically impossible to extend this state as a Hadamard state to the boundary of the domain of dependence of $S$. 

The relevance of the conjecture stems from the agreement that Hadamard states form the class of physical states for linear and perturbatively interacting quantum fields. For example, a failure of the Hadamard condition produces unbounded quantum fluctuations for certain observables in ultrastatic slab spacetimes \cite[Theorem 2.3]{Fewster-Verch}.

A subtle point is that the failure of the Hadamard condition could be sufficiently mild, such that the renormalised stress-energy tensor and its fluctuations remain well-defined. We thus examine in detail the black hole evaporation setting in Sec. \ref{sec:InformationLoss}, and conclude that the most likely semiclassical scenario is that of the development of a final curvature singularity. This motivates a more general {\it quantum very strong cosmic censorship} for semiclassical gravity, stated in Sec. \ref{sec:InformationLoss}. Finally, we devote Sec. \ref{sec:QG} to speculations on the role of quantum gravity in black hole evaporation as a unitary process.

\section{Quantum strong cosmic censorship}
\label{sec:QSCC}

In this section, we state a precise quantum strong cosmic censorship conjecture. We explain the role of the hypotheses in the conjecture and present evidence for its reasonableness. 

Throughout this paper a spacetime $(M, g_{ab})$ is a connected, paracompact, Hausdorff, time-orientable, Lorentzian, smooth manifold of spacetime dimension $n \geq 4$. The definition of the domain of dependence of a closed achronal set is used with respect to causal (not only timelike) curves, and we always consider achronal sets of co-dimension 1. We use units with $\hbar = c = 1$ and abstract index notation for tensors.

\subsection{The quantum strong cosmic censorship conjecture}

\begin{defn}
Let $(M, g_{ab})$ be a spacetime and $S$ a partial Cauchy surface, i.e., a closed achronal set with $D(S) \neq J(S)$. We say that $S$ is a {\bf strictly partial Cauchy surface} of $(M, g_{ab})$ if {\bf (a)} $S$ cannot be extended to be a Cauchy surface of the conformal completion of $(M, g_{ab})$ or {\bf (b)} if $S$ admits any extension to a Cauchy surface in the conformal completion of $(M, g_{ab})$, say $C$, the extension is such that $C\setminus S$ necessarily contains an open set of $C$ in the interior of the conformal completion of $(M, g_{ab})$.
\end{defn}

\begin{defn}
We say that the $\star$-algebra $\hat{\mathscr{A}}$ is an {\bf algebra extension} of the $\star$-algebra $\mathscr{A}$ if $\mathscr{A}$ is a $\star$-subalgebra $\hat{\mathscr{A}}$. Let $\omega: \mathscr{A} \to \mathbb{C}$ be an algebraic state. We say $\hat \omega: \hat{\mathscr{A}} \to \mathbb{C}$ is a {\bf state extension} of $\omega$ if $\hat \omega(a) = \omega(a)$ for every $a \in \mathscr{A}$.
\end{defn}

\begin{conj}[Quantum strong cosmic censorship]
Let $S$ be strictly partial Cauchy surface of the (not necessarily globally hyperbolic) inextendible spacetime $(M, g_{ab})$ and let $D(S)$ be its domain of dependence. $(D(S), \hat g_{ab})$ can be seen as a globally hyperbolic spacetime in its own right, where $\hat g_{ab} = \psi^{-1*} g_{ab}$ with $\psi: D(S) \to \psi(D(S)) \subset{M}$ an isometric embedding. Let $\Af$ be an $F$-local free (or perturbatively interacting) quantum field theory defined over $(M, g_{ab})$ and $\Bf$ a free (or perturbatively interacting) quantum field theory over $(D(S), \hat{g}_{ab})$ isomorphic to $\Af(M;D(S))$. Let $\omega: \Bf \to \mathbb{C}$ be a generic pure Hadamard state. Then, generically there exist no extension of $\omega$ to a state $\overline{\omega}: \Af(M; \overline{D(S)}) \to \mathbb{C}$ that is Hadamard.
\label{C1}
\end{conj}

The following remarks are in place:

\begin{rem}
On the hypothesis of $F$-locality: The failure of the algebra $\Af$ to be $F$-local, in the sense of Kay \cite{Floc}, might bring in the consequence that there exists a point $p \in D(S)$ for which there is no globally hyperbolic neighbourhood, $N_p$, such that $\Bf(D(S);N_p)$ is isomorphic to $\Af(M;N_p)$. A particular instance of this is if one can find an algebra element $b \in \Bf$ that is not isomorphic to any element of $\Af(M;N_p)$. In this case, it is clear that there is no state $\overline{\omega}: \Af(M; \overline{D(S)}) \to \mathbb{C}$ that can be an extension of any $\omega: \Bf \to \mathbb{C}$ (pure or otherwise), and the conjecture is meaningless.

However, note that a sensible conjecture without the $F$-locality hypothesis can be stated as follows:

\begin{conj}[Quantum strong cosmic censorship without $F$-locality]
Let $S$ be strictly partial Cauchy surface of the (not necessarily globally hyperbolic) inextendible spacetime $(M, g_{ab})$ and let $D(S)$ be its domain of dependence. $(D(S), \hat g_{ab})$ can be seen as a globally hyperbolic spacetime in its own right, where $\hat g_{ab} = \psi^{-1*} g_{ab}$ with $\psi: D(S) \to \psi(D(S)) \subset{M}$ an isometric embedding. Let $\Bf$ be a free (or perturbatively interacting) quantum field theory over $(D(S), \hat{g}_{ab})$ and $\omega: \Bf \to \mathbb{C}$ be a generic pure Hadamard state with two-point function $\omega_2 \in \mathscr{D}'(D(S) \times D(S))$. Then, the extension of $\omega_2$ as a distribution in $\mathscr{D}'(\overline{D(S)} \times \overline{D(S)})$ generically fails to satisfy the Hadamard condition at the boundary.
\label{C1-NoFLoc}
\end{conj}

\end{rem}

\begin{rem}
On the hypothesis of purity: There are clearly mixed Hadamard states that admit extensions as Hadamard states. For example, a state at the Unruh temperature in the Rindler wedge defined in the region $x > |t|$ of Minkowski spacetime with the induced metric, extends to the Minkowski vacuum. Generically, pure states in globally hyperbolic domains restrict to mixed states in a smaller domain.
\end{rem}

\begin{rem}
On the strict partial Cauchy surface hypothesis: 
Consider the two-dimensional Milne spacetime, $(F, g_{ab})$, the future wedge of Minkowski spacetime $t > |x|$ with the induced flat metric,
\begin{align}
g_{ab}^F = - \ee^{2 a \tau} (\dd \tau_a \dd \tau_b - \dd \chi_a \dd \chi_b) ,
\end{align}
here written in intrinsic future Rindler coordinates with $t = a^{-1} \ee^{a \tau} \cosh (a \chi)$, $x = a^{-1} \ee^{a \tau} \sinh (a \chi)$ (with $\tau, \chi \in \mathbb{R}$ and $a$ a positive constant). Consider now the Fock quantisation of a scalar with Fock vacuum vector $\Omega_{F}$ and the field formally represented as \cite{Birrell-Davies}
\begin{align}
\hat \Phi(\x) & = \int_\mathbb{R} \dd k \left(u_k(\x) \hat a_k + \overline{u_k}(\x) \hat a^\star_k\right), \\ u_k(\x) & = \left(\frac{\ee^{\pi k/a}}{8a}\right)^{1/2} \ee^{\ii k \chi}H^{(2)}_{\ii k/a}(m\ee^{a \tau}/a), \label{MilneModes}
\end{align}
where $H^{(2)}_\nu$ is a Hankel function of the second kind. The modes \eqref{MilneModes} are in fact a superposition of positive-frequency Minkowski modes \cite{FPH},
\begin{align}
u_k(\x) = (8 \pi^2 a)^{-1/2} \int_\mathbb{R} \dd \rho \ee^{-\ii \omega(\rho) t} \ee^{\ii p(\rho) x} \ee^{-\ii k \rho/a},
\end{align}
with $p(\rho) = - m \sinh \rho$ and $\omega(\rho) = (p^2 + m^2)^{1/2}$, and the two-point function can be written as
\begin{align}
\omega_2^F(\x, \x') = (8 \pi^2 a)^{-1} \int_{\mathbb{R}^3} \dd k \dd \rho \dd \rho' \ee^{-\ii k(\rho-\rho')/a} \ee^{-\ii \omega(\rho) t + \ii p(\rho) x} \ee^{\ii \omega(\rho') t' - \ii p(\rho') x'}. 
\end{align}

Carrying out the Fourier transform in $k$ one can obtain a term proportional to $\delta(\rho - \rho')$, and then trivially perform a second integral. The remaining integral, under the change of variables $p = - m \sinh \rho$, can be seen to yield
\begin{align}
\omega_2^F(\x, \x') =  \int_{\mathbb{R}} \frac{\dd p}{4 \pi \omega_p} \ee^{-\ii \omega_p(t-t') + \ii p (x-x')},
\end{align}
which is nothing but the Minkowski vacuum restricted to $(F, g_{ab}^F)$. Thus, obviously $\omega_2^F$ admits an extension to the whole of Minkowski spacetime as a Hadamard state. However, any Cauchy surface of the Milne universe will fail to be a strictly partial Cauchy surface in Minkowski spacetime.

The former example is clearly connected the well-known `hyperboloid information loss' example of Wald. Clearly, the state $\omega_F$ and the Minkowski vacuum are unitarily inequivalent, even if the Bogoliubov $\beta$-coefficients vanish. This is so because the inner products of the two quantum theories are not equivalent. To wit, it is clear that any data compactly supported in the $|x| > t$ portion of future null infinity does not enter the future hyperboloid. As a result, it is impossible to uniquely evolve $\omega_F$ towards the past and obtain the Minkowski vacuum. However, the Minkowski vacuum is the complex analytic extension of $\omega_F$ to Minkowski spacetime (which is also the analytic extension of the Milne universe). Thus, when restricting only to analytic data in Minkowski space, no information is lost on the hyperboloid, as such data can be recovered in $\mathscr{I}^+$ by analyticity. It is this very special feature that seems to avoid any breakdown of the Hadamard property on the past Cauchy horizon, which is however not expected to occur generically.

\end{rem}

\begin{rem}
\label{rem:Conf}
On the genericness  of the conjecture: It is possible to finely tune Hadamard states that can be extended beyond Cauchy horizons. Consider a globally hyperbolic spacetimes $(M, g_{ab})$ and a Hadamard state, $\omega$ defined on a conformal quantum field theory in the whole spacetime. Consider now a scalar function $\Omega: M \to \mathbb{R}^+$ that blows up as the point $p \in M$ is approached and is smooth in $M \backslash \{ p \}$. The new spacetime $(M\backslash \{ p \}, \Omega^2 g_{ab})$ is inextendible and contains a Cauchy horizon corresponding to where the future lightcone of $p$ lies in the original spacetime. However, the conformal state $\omega^\Omega$ is Hadamard throughout the new spacetime. 
\end{rem}

It is important that the spacetime $(M, g_{ab})$ be inextendible, both in the above conjectures and in discussions of Penrose's strong cosmic censorship more generally. In extendible spacetimes, one can easily find counter-examples, as shown in the appendix.

\subsection{Supporting evidence}

We now discuss some evidence in support of the reasonableness of conjecture \ref{C1}:

\begin{enumerate}
\item It is well-known that, in spacetimes with bifurcate Killing horizons, states in the Fock space construction of the wedge vacua (e.g. Boulware or Fulling-Rindler vacua) fail to be Hadamard at the future and past Cauchy horizons of the wedges, which lie along the bifurcate Killing horizon.
\item In \cite{Juarez-Aubry:2021}, it is shown that in generalised Reissner-Nordstr{\"o}m-like black holes in two spacetime dimensions, the Hadamard condition is lost for the HHI and Unruh vacua at Cauchy horizons, as probed by particle detectors and by the stress-energy flux. These results had been anticipated in \cite{Juarez-Aubry:2015}. 
\item The same conclusion as 2. holds in $3+1$ Reissner-Nordtr\"om \cite{Lanir:2018, Zilberman:2019}.
\item The Reissner-Nordstr\"om de Sitter case has been studied in \cite{Hollands:2019, Hollands:2020} and in \cite{Papadodimas:2019msp, Shrivastava:2020xmw}. States that are regular at the cosmological horizon generically fail to be Hadamard at the inner horizon, where the expectation value of the stress-energy tensor diverges more strongly than the classical stress-energy tensor, see \cite{Cardoso:2017soq, Dias:2018ufh}.
\item  In \cite{Zilberman:2022aum} it is shown that the renormalised stress-energy flux in the Unruh state in Kerr spacetime diverges at the Cauchy inner horizon.
\item In \cite{Dias:2019ery} the HHI state for a massless scalar field in the near-extremal BTZ black hole has been studied. The authors find that the HHI state is not Hadamard at the BTZ inner horizon, but the expectation value of the stress-energy tensor is finite. However, in \cite{Emparan:2020rnp} it is argued that the solution should not be stable under semiclassical back-reaction.
\item In \cite{McMaken:2023uue} it is shown that a class of ``inner-extremal" regular black holes, which avoid mass-inflation classical instabilities, have diverging stress-energy tensor in the Unruh vacuum at the Cauchy horizon. 
\end{enumerate}

In addition to the former, there are known structural results on the breakdown of the Hadamard property at Cauchy horizons in certain cases.

\begin{itemize}
\item The KRW theorems \cite{KRW} establish that there are no Hadamard state extensions for a free scalar field in a globally hyperbolic region with a compactly generated Cauchy horizons. In fact, the theorems are stronger in the following sense: Theorem KRW 1 establishes that there is no $F$-local extension of an initial globally hyperbolic quantum field theory at the Cauchy horizon, due to a breakdown of $F$-locality at certain \emph{base points} of the Cauchy horizon. Theorem KRW 2 establishes that at these base points a state extension will fail to be Hadamard in a very strong sense: it is not possible that the difference between the state's two-point bi-distribution and a locally constructed Hadamard bi-distribution be a bounded function in a neighbourhood of any base point in $M \times M$.
\item Prop. 3.2 in \cite{Fewster-Verch} implies that any state that is pure or normal to a pure state (i.e., represented as a density operator in the Hilbert space of the pure state) defined in the interior of a double cone in a globally hyperbolic spacetime cannot be extended to the spacelike boundary of the double cone as a Hadamard state. Note here that the boundary of the double cone plays the role of a Cauchy horizon for the interior region.
\end{itemize}

\section{Implications for the information loss puzzle}
\label{sec:InformationLoss}

\begin{figure}
\begin{center}
\begin{tikzpicture}
\draw[thick] (-2,-2) -- (-2, 2);
\draw[thick] (-2,-2) -- (1.75, 1.75);
\draw[decoration = {zigzag,segment length = 2mm, amplitude = 0.5mm}, decorate] (-2,2) -- (-0.5,2);
\draw[decoration = {zigzag,segment length = 2mm, amplitude = 0.5mm}, decorate] (-0.5,2) -- (0.5,3); 
\draw [thick, draw=none, fill=blue, fill opacity=0.2]
       (-2,2) -- (-0.5,2) -- (-2,0.5) -- cycle; 
\draw [thick, draw=none, pattern=north west lines, pattern color = black]
       (-2,2) -- (-1.2,2) to[out=-65,in=55] (-2,-2) -- cycle;  
\draw[thick]  (-1.2,2) to[out=-65,in=55] (-2,-2); 
\draw[thick] (0.5, 3) -- (1.75,1.75);
\draw[thick] (-0.5,2) -- (-2, 0.5);
\node at (-1.8,-2.1) {$i^-$};
\node at (0.7, 3.2) {$i^+$};
\node at (2,1.75) {$i^0$};
\node at (1.4,2.6) {$\mathscr{I}^+$};
\node at (0.5,-0) {$\mathscr{I}^-$};
\node at (-2.5, 0) {$r = 0$};
\node at (-1.2, 2.2) {$r \to 0$};
\end{tikzpicture}
\end{center}
\caption{Breakdown of semiclassical gravity in black hole evaporation: Infalling matter forms a black hole region (blue). At the end-point of evaporation, a semiclassical final singularity is formed, replacing the Cauchy horizon in Fig. \ref{Fig:Hawking-Diagram}. The final singularity stretches all the way to $\mathscr{I}^+$.}  
\label{Fig:Conjecture}
\end{figure}
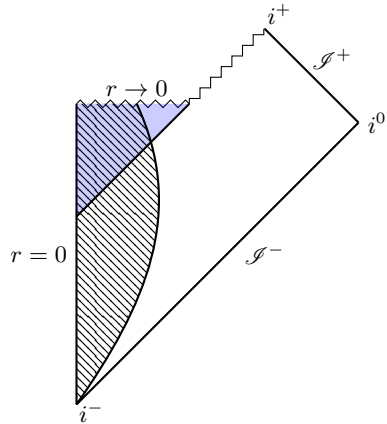

A breakdown of the Hadamard condition in the Cauchy horizon of Fig. \ref{Fig:Hawking-Diagram} would pose serious questions on the physicality of the evaporation diagram. In this section, we argue that the semiclassical physics picture of black hole evaporation should be replaced by one in which a final spacetime singularity is developed instead of a post-evaporation region, cf. Fig. \ref{Fig:Conjecture}.\footnote{We assume, like Hawking, that in semiclassical gravity one actually has black hole formation with a singularity at the origin. For other views, see \cite{
Baccetti:2017oas}, but see also \cite{Juarez-Aubry:2018, Arderucio-Costa:2017etb} in favour of our assumption.}

Let us begin by a cautionary remark. The reader might wonder if the KRW theorems do not already guarantee that the Hadamard condition is lost at the Cauchy horizon of Fig. \ref{Fig:Hawking-Diagram}. After all, the Cauchy horizon seems to emanate from the singularity at $r = 0$, and it is therefore reasonable to believe that it is compactly generated. One could for example consider a compact set defined by a closed finite-radius ball centered around the final evaporation point. The issue is however that the evaporation event is not a spacetime point. This has been strongly emphasised e.g. in \cite{Maudlin}. Thus, in order to apply the KRW theorems one needs to remove final evaporation event. But now the putative ball is no longer compact. So it seems that the situation is, due to a topological subtlety, outside of the scope of the KRW theorems.\footnote{The author thanks Bernard Kay for providing this argument in an email exchange.} One could argue in favour of removing a small open set around the evaporation event, for example, by assuming that the breakdown of a spacetime description of gravity occurs already in an open neighbourhood around the singular evaporation event. Proceeding in this way, however, we cannot argue that semiclassical gravity breaks down, as this now becomes an assumption.

The evaporation event however allows one to put forth a strong argument suggesting the breakdown of semiclassical theory at the Cauchy horizon of Fig. \ref{Fig:Hawking-Diagram}. On the one hand, it is well-known that the Hadamard property of states breaks down at curvature singularities, see e.g. \cite{Juarez-Aubry:2014jba}.\footnote{Note that black hole singularities generically contain blow-ups in the Weyl tensor, unlike in the example of remark~\ref{rem:Conf} above.} Thus, the Hadamard property must break down at the final evaporation event. On the other hand, this final singularity must propagate in the geometric side of the semiclassical Einstein equations, such that, if the semiclassical gravity equations are well posed, it does so along null geodesics \cite{Hormander}.\footnote{Here, we are assuming that the geometric sector of semiclassical gravity can be described by a hyperbolic system. The claim then follows from the non-linear version of the propagation of singularities theorem \cite{Hormander}.} This gives support, not only the breakdown of the Hadamard property, but also to the development of a curvature singularity along the horizon.

The semiclassical picture that goes along with this mathematical statement is as follows: Towards the end of evaporation the mass of the black hole tends to zero and the Hawking temperature becomes unboundedly large, with the state's two-point function developing new divergences. To see that this must be the case, for the sake of illustration, consider a massless scalar in a KMS state at $T = 1/\beta>0$ in two-dimensional Minkowski spacetime. Modulo infrared ambiguities, its two point function is
\begin{align}
\omega_\beta(\Phi(\x)\Phi(\x')) = -\frac{1}{4 \pi} \left( \ln \sinh(\pi(\Delta x + \Delta t - \ii \epsilon)/\beta) + \ln \sinh(\pi(\Delta x - \Delta t + \ii \epsilon)/\beta) \right).
\end{align}

To \emph{unambiguously} probe the Hadamard structure as $\beta \to 0^+$, it suffices to inspect the expectation value of \emph{unambiguous renormalised observables}. 
For example, the stress-energy tensor,
\begin{align}
\omega_\beta(T_{ab}) = \frac{\pi}{3 \beta^2} \delta_{ab},
\end{align}
diverges polynomially as $\beta \to 0^+$, so $\omega_\beta$ cannot be Hadamard.

Thus, the final stage of black hole evaporation seems to be described semiclassically described by unboundedly hot Hawking radiation and a curvature singularity in place of the Cauchy horizon of Fig. \ref{Fig:Hawking-Diagram}, suggesting no information loss. See Fig. \ref{Fig:Conjecture}. We should mention that numerical evidence supporting this picture is available since the work of Hawking and Stewart \cite{Hawking:1992ti}.

The above discussion motivates us to conjecture that generically in semiclassical gravity $\omega$ cannot be extended to a state $\overline \omega$ that differs from a Hadamard state by a regular term at the Cauchy horizon. More precisely:

\begin{conj}[Quantum very strong cosmic censorship]
Let $(M, g_{ab})$ be an inextendible spacetime and $S$, $D(S)$, $\Af$, $\Bf$ and $\omega$ be as in conjecture \ref{C1}. Suppose furthermore that the pair $(D(S), \omega)$ satisfies the semiclassical gravity equations. Then, generically there exists no extension of $\omega$ to a state $\overline{\omega}: \Af(M; \overline{D(S)}) \to \mathbb{C}$ such that $\overline{\omega} - H \in L^1_{\rm loc}$ for $H$ a Hadamard parametrix on $\overline{D(S)}$, i.e., $\overline{\omega}-H$ cannot be defined distributionally on $\overline{D(S)}$.
\label{C2}
\end{conj}

It is clear how to relax the $F$-locality assumption from conjecture \ref{C2} in view of conjecture \ref{C1-NoFLoc}.

\section{The role of quantum gravity and final remarks}
\label{sec:QG}

It is a general expectation that quantum gravity should `cure' spacetime singularities. 
This is reasonable: Consider in particular curvature singularities, then as curvature invariants approximate the scale of suitable powers of inverse Planck's length, say e.g. $\ell_{\rm P}^2 R = O(1)$, we are presumably entering the regime of full quantum gravity, and not of classical geometry.

\begin{figure}
\begin{center}
\begin{tikzpicture}
\draw[thick] (-2,-2) -- (-2, 2);
\draw[thick] (-2,-2) -- (1.75, 1.75);
\draw [thick, draw=none, fill=blue, fill opacity=0.2]
       (-2,2) -- (-0.5,2) -- (-2,0.5) -- cycle; 
\draw [thick, draw=none, pattern=north west lines, pattern color = black]
       (-2,5.5) -- (-1.2,2) to[out=-65,in=55] (-2,-2) -- cycle;  
\draw[thick]  (-1.2,2) to[out=-65,in=55] (-2,-2); 
\draw[thick] (-2,5.5) -- (-1.2, 2);
\draw[thick] (-0.5, 4) -- (1.75,1.75);
\draw[thick] (1.75, 1.75) -- (-2, 5.5);
\draw[thick] (-2, 2) -- (-2, 5.5);
\draw[thick] (-0.5,2) -- (-2, 0.5);
\draw [thick, draw=none, pattern=crosshatch dots, pattern color=red, fill opacity=0.5]
       (-2,2) -- (-2,5.5) -- (0.5,3) -- (-0.5,2)-- cycle; 
\draw[line width=2mm, rounded corners, color=black] {(-2,2) -- (-0.5,2) -- (0.5,3)};
\node at (-1.8,-2.1) {$i^-$};
\node at (-1.8,5.7) {$i^+$};
\node at (2,1.75) {$i^0$};
\node at (0.5,3.5) {$\mathscr{I}^+$};
\node at (0.5,0) {$\mathscr{I}^-$};
\node at (-2.5, 2.1) {$r = 0$};
\end{tikzpicture}
\end{center}
\caption{A quantum gravity `region', denoted as solid black, separates a pre-evaporation region, with would-be black-hole region shaded in blue, from a post-evaporation region, denoted by red dots. The evaporation diagram can be described, modulo the quantum gravity `region', as a globally hyperbolic spacetime.}  
\label{Fig:QG}
\end{figure}
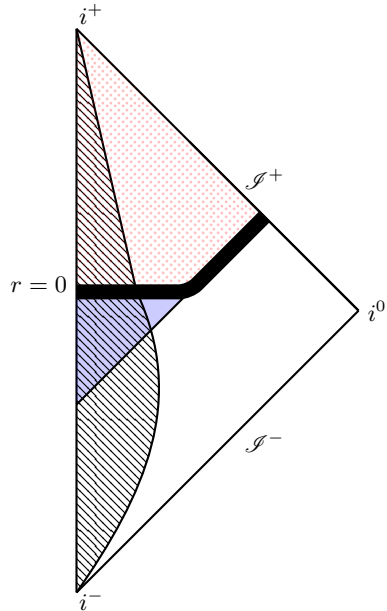

Here we advocate that full quantum gravity will kick in towards the end of evaporation and cure the semiclassically predicted final singularity in Fig. \ref{Fig:Conjecture}.  This represents a radical departure from the narrative advocated in many quantum gravity speculations, whereby the `quantum gravity region' in evaporating diagrams is confined to a (say, small) `compact neighbourhood' of the black hole origin singularity. See for example the diagrams presented in \cite{Hossenfelder:2009xq, Bianchi:2018mml, Ashtekar:2005cj, Ashtekar:2008jd}.
This view is represented in the spherically-symmetric case in Fig. \ref{Fig:QG}, where we emphasise that, modulo this quantum-gravity-described region, the resulting `spacetime' has globally hyperbolic `features' and information is preserved. The details of the post-evaporation region in the diagram are highly speculative. For example, we do not know how the collapsing matter and quantum gravity interact in the full quantum gravity regime. For concreteness, we have assumed that some matter `goes through' the otherwise-black-hole origin singularity and continues to behave largely classically at late times -- so the spacetime in the post-evaporation looks as sourced by a spherically symmetric distribution of matter and filled with radiation. However, the details of the post-evaporation region are not essential for the full picture that we advocate or the nature of the quantum gravity `region' in the evaporating diagram in Fig. \ref{Fig:QG}.

To wrap up the discussion, in 1992 Preskill wrote an influential note asking whether black holes destroy information \cite{Preskill:1992tc}. In our conservative view,\footnote{\emph{Conservative} in the sense of \cite{Hossenfelder:2009xq}.} semiclassical black holes do not destroy information any differently to how classical black holes destroy information in the interior region, with the addition that the former also destroy information in the exterior black hole region at very late times. However, there is no breakdown of classical determinism, information loss or hints that quantum gravity should be non-unitary.

\section*{Acknowledgments}
The author thanks especially BS Kay for his interest in our conjectures and strong encouragement to write this manuscript. Warm thanks are due to B Arderucio Costa for many active discussions on the topic. The author also thanks J Louko for prompting the study of the Milne universe example, which led the author to imposing the strictly partial Cauchy condition in the conjectures, and CJ Fewster for pointing out the relevance of ref. \cite{Fewster-Verch}. Thanks are also due to the participants 2022 Black Hole Information Puzzle Workshop at the John Bell Institute for very stimulating discussions on the topic. The author thanks anonymous criticism that pointed to finely-tuned conformal states as counter-examples to a previous, stronger version of Conjecture \ref{C1}. The author is currently supported by EPSRC, through the EPSRC Open Fellowship EP/Y014510/1. This work has been supported by a CONAHCYT (formerly CONACYT) Postdoctoral Research Fellowship and CONAHCYT FORDECYT-PRONACES grant 140630. 

\appendix

\section{``Violations" of strong cosmic censorship in extendible spacetimes}
\label{App}

One can see that Penrose's strong cosmic censorship is generically ``violated" in extendible spacetimes, and the situation is not different for quantum strong cosmic censorship by an analogous argument. The use of quotation marks above answers to the fact that strong cosmic censorship should not be discussed in the context of extendible spacetimes in the first place. We show the following results:

\begin{prop}
Penrose's strong cosmic censorship is generically ``violated" in extendible spacetimes.
\end{prop}
\begin{proof}
Let $(M, g_{ab})$ be a globally hyperbolic spacetime. Choose a compact set, $K$, and a Cauchy surface, $S$, such that the closure of $K$ is contained in the future of $S$. Consider now the spacetime $(\psi^* M \backslash \overline{K}, \psi^* g_{ab})$, where $\psi$ is an isometric embedding onto $(M, g_{ab})$. This new spacetime is extendible by construction. Furthermore, it is not globally hyperbolic and has a Cauchy horizon (with respect to $S$) at the (pullback of the) boundary of $J^+(K)$. However, $\psi^* M \backslash \overline{K}$ carries a smooth structure (since $\overline{K}$ is closed), so strong cosmic censorship is ``violated" if $(M, g_{ab})$ is a solution to the Einstein field equations.
\end{proof}

The analogous result in quantum strong cosmic censorship is:

\begin{prop}
Quantum strong cosmic censorship is generically ``violated" in extendible spacetimes.
\end{prop} 
\begin{proof}
Let $(M, g_{ab})$ be a globally hyperbolic spacetime. Choose a compact set $K$ and a Cauchy surface $S$, as above. Define a pure, Hadamard state $\omega$ in a neighbourhood $N$ of $S$ for, say, the free Klein-Gordon theory, such that $N$ does not intersect the closure of $K$. By the Fulling-Sweeny-Wald theorem \cite{Fulling:1978ht} this state is Hadamard everywhere in $(M, g_{ab})$. Consider now the spacetime $(\psi^* M \backslash \overline{K}, \psi^* g_{ab})$, where $\psi$ is an isometric embedding as before. The new spacetime is extendible, smooth, not globally hyperbolic and has a Cauchy horizon (with respect to $S$) at the (pullback of the) boundary of $J^+(K)$. The pullback of $\omega$ defines a locally Hadamard state everywhere in $(\psi^* M \backslash \overline{K}, \psi^* g_{ab})$, which is furthermore initially pure, since $\psi^* N \subset \psi^* M \backslash \overline{K}$.
\end{proof}


\begin{thebibliography}{99}

\bibitem{Hawking:1975}
S.~W.~Hawking,
Commun. Math. Phys. \textbf{43} (1975), 199-220
[erratum: Commun. Math. Phys. \textbf{46} (1976), 206]
doi:10.1007/BF02345020

\bibitem{Juarez-Aubry:2018}
B.~A.~Ju\'arez-Aubry and J.~Louko,
JHEP \textbf{05} (2018), 140
doi:10.1007/JHEP05(2018)140
[arXiv:1804.01228 [gr-qc]].

\bibitem{Kay:2022wpn}
B.~S.~Kay,
(2022) 
[arXiv:2206.07445 [hep-th]].

\bibitem{Unruh-Wald} 
W.~G.~Unruh and R.~M.~Wald,
Rept. Prog. Phys. \textbf{80} (2017) no.9, 092002
doi:10.1088/1361-6633/aa778e
[arXiv:1703.02140 [hep-th]].

\bibitem{Hossenfelder:2009xq}
S.~Hossenfelder and L.~Smolin,
Phys. Rev. D \textbf{81} (2010), 064009
doi:10.1103/PhysRevD.81.064009
[arXiv:0901.3156 [gr-qc]].

\bibitem{Preskill:1992tc} 
J.~Preskill,
``Do black holes destroy information?'', in {\it International Symposium on Black holes, Membranes, Wormholes and Superstrings} 22-39 (1992).
[arXiv:hep-th/9209058 [hep-th]].

\bibitem{Fewster-Verch}
C.~J.~Fewster and R.~Verch,
Class. Quant. Grav. \textbf{30} (2013), 235027
doi:10.1088/0264-9381/30/23/235027
[arXiv:1307.5242 [gr-qc]].

\bibitem{Floc}
B.~S.~Kay, 
Rev. Math. Phys. {\bf 4} (1992), 167-95
doi:10.1142/S0129055X92000194
(available at https://www.researchgate.net/publication/234525000\_The\_Principle\_of\_Locality\_and\_Quantum\_Field\_Theory\_on\_
non\_Globally\_Hyperbolic\_Curved\_Spacetimes.)


\bibitem{Birrell-Davies}
N.~D.~Birrell and P.~C.~W.~Davies, 
{\it Quantum Fields in curved space}
(Cambridge University Press, 1982).

\bibitem{FPH}
S.~A.~Fulling, L.~Parker and B.~L.~Hu,
Phys. Rev. D \textbf{10} (1974), 3905-3924
doi:10.1103/PhysRevD.10.3905;
erratum \textit{ibid} {\bf 11}, 1714.

\bibitem{Juarez-Aubry:2021}
B.~A.~Ju\'arez-Aubry and J.~Louko,
AVS Quantum Sci. \textbf{4} (2022) no.1, 013201
doi:10.1116/5.0073373
[arXiv:2109.14601 [gr-qc]].

\bibitem{Juarez-Aubry:2015}
B.~A.~Ju\'arez-Aubry,
Int. J. Mod. Phys. D \textbf{24} (2015) no.09, 1542005
doi:10.1142/S0218271815420055
[arXiv:1502.02533 [gr-qc]].

\bibitem{Lanir:2018}
A.~Lanir, A.~Ori, N.~Zilberman, O.~Sela, A.~Maline and A.~Levi,
Phys. Rev. D \textbf{99} (2019) no.6, 061502
doi:10.1103/PhysRevD.99.061502
[arXiv:1811.03672 [gr-qc]].

\bibitem{Zilberman:2019}
N.~Zilberman, A.~Levi and A.~Ori,
Phys. Rev. Lett. \textbf{124} (2020) no.17, 171302
doi:10.1103/PhysRevLett.124.171302
[arXiv:1906.11303 [gr-qc]].

\bibitem{Hollands:2019}
S.~Hollands, R.~M.~Wald and J.~Zahn,
Class. Quant. Grav. \textbf{37} (2020) no.11, 115009
doi:10.1088/1361-6382/ab8052
[arXiv:1912.06047 [gr-qc]].

\bibitem{Hollands:2020}
S.~Hollands, C.~Klein and J.~Zahn,
Phys. Rev. D \textbf{102} (2020) no.8, 085004
doi:10.1103/PhysRevD.102.085004
[arXiv:2006.10991 [gr-qc]].

\bibitem{Papadodimas:2019msp}
K.~Papadodimas, S.~Raju and P.~Shrivastava,
JHEP \textbf{12} (2020), 003
doi:10.1007/JHEP12(2020)003
[arXiv:1910.02992 [hep-th]].

\bibitem{Shrivastava:2020xmw}
P.~Shrivastava,
[arXiv:2009.03261 [hep-th]].

\bibitem{Cardoso:2017soq}
V.~Cardoso, J.~L.~Costa, K.~Destounis, P.~Hintz and A.~Jansen,
Phys. Rev. Lett. \textbf{120} (2018) no.3, 031103
doi:10.1103/PhysRevLett.120.031103
[arXiv:1711.10502 [gr-qc]].

\bibitem{Dias:2018ufh}
O.~J.~C.~Dias, H.~S.~Reall and J.~E.~Santos,
Class. Quant. Grav. \textbf{36} (2019) no.4, 045005
doi:10.1088/1361-6382/aafcf2
[arXiv:1808.04832 [gr-qc]].

\bibitem{Zilberman:2022aum}
N.~Zilberman, M.~Casals, A.~Ori and A.~C.~Ottewill,
Phys. Rev. Lett. \textbf{129} (2022) no.26, 261102
doi:10.1103/PhysRevLett.129.261102
[arXiv:2203.08502 [gr-qc]].

\bibitem{Dias:2019ery}
O.~J.~C.~Dias, H.~S.~Reall and J.~E.~Santos,
JHEP \textbf{12} (2019), 097
doi:10.1007/JHEP12(2019)097
[arXiv:1906.08265 [hep-th]].

\bibitem{Emparan:2020rnp}
R.~Emparan and M.~Toma\v{s}evi\'c,
JHEP \textbf{06} (2020), 038
doi:10.1007/JHEP06(2020)038
[arXiv:2002.02083 [hep-th]].

\bibitem{McMaken:2023uue}
T.~McMaken,
Phys. Rev. D \textbf{107} (2023) no.12, 125023
doi:10.1103/PhysRevD.107.125023
[arXiv:2303.03562 [gr-qc]].

\bibitem{KRW}
B.~S.~Kay, M.~J.~Radzikowski and R.~M.~Wald,
Commun. Math. Phys. \textbf{183} (1997), 533-556
doi:10.1007/s002200050042
[arXiv:gr-qc/9603012 [gr-qc]].

\bibitem{Baccetti:2017oas}
V.~Baccetti, R.~B.~Mann and D.~R.~Terno,
Int. J. Mod. Phys. D \textbf{26} (2017) no.12, 1743008
doi:10.1142/S0218271817170088
[arXiv:1706.01180 [gr-qc]].

\bibitem{Arderucio-Costa:2017etb}
B.~Arderucio-Costa and W.~Unruh,
Phys. Rev. D \textbf{97} (2018) no.2, 024005
doi:10.1103/PhysRevD.97.024005
[arXiv:1709.00115 [gr-qc]].

\bibitem{Maudlin}
T.~Maudlin,
[arXiv:1705.03541 [physics.hist-ph]].

\bibitem{Juarez-Aubry:2014jba}
B.~A.~Ju\'arez-Aubry and J.~Louko,
Class. Quant. Grav. \textbf{31} (2014) no.24, 245007
doi:10.1088/0264-9381/31/24/245007
[arXiv:1406.2574 [gr-qc]].

\bibitem{Hormander}
L.~H\"ormander, {\it Lectures on nonlinear hyperbolic differential equations} (Springer, 2003).

\bibitem{Hawking:1992ti}
S.~W.~Hawking and J.~M.~Stewart,
Nucl. Phys. B \textbf{400} (1993), 393-415
doi:10.1016/0550-3213(93)90410-Q
[arXiv:hep-th/9207105 [hep-th]].

\bibitem{Bianchi:2018mml}
E.~Bianchi, M.~Christodoulou, F.~D'Ambrosio, H.~M.~Haggard and C.~Rovelli,
Class. Quant. Grav. \textbf{35} (2018) no.22, 225003
doi:10.1088/1361-6382/aae550
[arXiv:1802.04264 [gr-qc]].


\bibitem{Ashtekar:2005cj}
A.~Ashtekar and M.~Bojowald,
Class. Quant. Grav. \textbf{22} (2005), 3349-3362
doi:10.1088/0264-9381/22/16/014
[arXiv:gr-qc/0504029 [gr-qc]].

\bibitem{Ashtekar:2008jd}
A.~Ashtekar, V.~Taveras and M.~Varadarajan,
Phys. Rev. Lett. \textbf{100} (2008), 211302
doi:10.1103/PhysRevLett.100.211302
[arXiv:0801.1811 [gr-qc]].

\bibitem{Fulling:1978ht}
S.~A.~Fulling, M.~Sweeny and R.~M.~Wald,
Commun. Math. Phys. \textbf{63} (1978), 257-264
doi:10.1007/BF01196934.

\end{thebibliography}
\end{document}